\documentclass{ws-ijmpa}

\newcommand{\be}{\begin{equation}}
\newcommand{\ee}{\end{equation}}
\newcommand{\bea}{\begin{eqnarray}}
\newcommand{\eea}{\end{eqnarray}}
\newcommand{\bean}{\begin{eqnarray*}}
\newcommand{\eean}{\end{eqnarray*}}
\newcommand{\ba}{\begin{array}}
\newcommand{\ea}{\end{array}}

\newcommand{\slashl}[1]{\not{\!\!#1}}
\newcommand{\slashs}[1]{\not{\!#1}}

\begin{document}

%

\def\nocropmarks{\vskip5pt\phantom{cropmarks}}


%

\markboth{H. Kawamura, J. Kodaira, C.F. Qiao and K. Tanaka}
{$B$ Meson Light-cone Distribution Amplitudes and Heavy-quark Symmetry}

%
\catchline{}{}{}
%

\setcounter{page}{1}

\title{$B$ MESON LIGHT-CONE DISTRIBUTION AMPLITUDES\\
AND HEAVY-QUARK SYMMETRY\footnote{Talk presented by K. Tanaka.}
}

\author{\footnotesize HIROYUKI KAWAMURA
}

\address{Deutsches Elektronen-Synchrotron, DESY\\
Platanenallee 6, D 15738 Zeuthen, GERMANY
}

\author{JIRO KODAIRA\, and CONG-FENG QIAO\footnote{JSPS Research Fellow.}}

\address{Dept. of Physics, Hiroshima University\\
Higashi-Hiroshima 739-8526, JAPAN
}

\author{KAZUHIRO TANAKA}

\address{Dept. of Physics, Juntendo University\\
Inba-gun, Chiba 270-1695, JAPAN
}

\maketitle


\begin{abstract}
We present a systematic study of the $B$ meson light-cone distribution amplitudes
which are relevant for the QCD factorization
approach for the exclusive $B$ meson decays. 
We construct representations for the
quark-antiquark distribution amplitudes in terms of independent 
dynamical degrees of freedom,
which exactly satisfy the QCD equations of motion 
and constraints from heavy-quark symmetry.
\end{abstract}

\vspace{0.7cm}
\noindent
Recently systematic methods based on the QCD factorization
have been developed for the exclusive $B$ meson decays
into light mesons\cite{Beneke:2000ry,Beneke:2001wa,Beneke:2001at,Bauer:2001cu}
(for other approaches see\cite{Ball:1998kk}).
Essential ingredients in this approach are
the light-cone distribution amplitudes for the participating mesons,
which constitute nonperturbative long-distance contribution
to the factorized amplitudes.
{}For the light mesons ($\pi$, $K$, 
$\rho$, 
$K^{*}$, etc.)
appearing in the final state,
systematic model-independent study of the light-cone distributions 
exists for both leading 
and higher twists.\cite{Braun:1990iv}
On the other hand, 
the light-cone distribution amplitudes for
the $B$ meson are not well-known at present and 
provide a major source of uncertainty in the calculations of 
the decay rates.
In this work,\cite{KKQT} 
we demonstrate that 
heavy-quark symmetry and constraints from the equations of motion
determine a unique analytic solution for the $B$ meson light-cone
distribution amplitudes within the two-particle Fock states.
We also derive the exact integral representations 
for the effects of higher Fock states with additional gluons.

In the heavy-quark limit, the $B$ meson matrix elements obey 
the heavy-quark symmetry,
and is described by the heavy-quark
effective theory (HQET).\cite{Neubert:1994mb}
Following Refs.\cite{Grozin:1997pq,Beneke:2001wa},
we introduce the quark-antiquark light-cone distribution 
amplitudes $\tilde{\phi}_{\pm}(t)$ of the $B$ meson in terms of matrix element
in the HQET:
\be
\langle 0 | \bar{q}(z) \Gamma h_{v}(0) |\bar{B}(p) \rangle
 = - \frac{i f_{B} M}{2} {\rm Tr}
 \left[ \gamma_{5}\Gamma \frac{1 + \slashs{v}}{2}
 \left\{ \tilde{\phi}_{+}(t) - \slashs{z} \frac{\tilde{\phi}_{+}(t)
 -\tilde{\phi}_{-}(t)}{2t}\right\} \right]\ . 
 \label{phi}
\ee
where $z_{\mu}$ is a light-like vector ($z^{2}=0$), $v^{2} = 1$,
$t=v\cdot z$,
and $p^{\mu} = Mv^{\mu}$
is the 4-momentum of the $B$ meson with mass $M$.
$h_{v}(x)$ denotes the effective $b$-quark 
field,\cite{Neubert:1994mb}
$b(x) \approx \exp(-im_{b} v\cdot x)h_{v}(x)$, 
and is subject to the on-shell 
constraint, $\slashl{v} h_{v} = h_{v}$.
$\Gamma$ is a generic Dirac matrix and,
here and in the following, the path-ordered gauge
factors are implied in between the constituent fields.
$f_{B}$ is the decay constant defined as usually as
%
$\langle 0 | \bar{q}(0) \gamma^{\mu}\gamma_{5} h_{v}(0) |\bar{B}(p) \rangle
   = i f_{B} M v^{\mu}$
%
so that $\tilde{\phi}_{\pm}(t=0) = 1$.
$\tilde{\phi}_{+}$ is of leading-twist,
whereas $\tilde{\phi}_{-}$ has subleading twist.\cite{Grozin:1997pq} 

It is well-known that the QCD equations of motion
impose a set of relations between distribution 
amplitudes for the light-mesons.\cite{Braun:1990iv}
The corresponding relations can be derived
from the identities between the nonlocal operators:
\bea
 \frac{\partial}{\partial x^{\mu}}
 \bar{q}(x) \gamma^{\mu} \Gamma h_{v}(0)
  &=& 
i \int_{0}^{1}duu \ \bar{q}(x) gG_{\mu \nu}(ux) x^{\nu}
  \gamma^{\mu}\Gamma h_{v}(0) \ ,
 \label{id1} \\
 v^{\mu}\frac{\partial}{\partial x^{\mu}}
 \bar{q}(x) \Gamma h_{v}(0)
 &=& 
i \int_{0}^{1}du (u-1)\ \bar{q}(x) gG_{\mu \nu}(ux) 
 v^{\mu}x^{\nu}\Gamma h_{v}(0) \nonumber \\
 &+& 
 v^{\mu}\left.
   \frac{\partial}{\partial y_{\mu}}\
  \bar{q}(x+y) \Gamma h_{v}(y)\right|_{y \rightarrow 0}\ ,
\label{id2}
\eea
%
where $G_{\mu \nu}$ is the gluon field strength tensor, and
we have used the equations of motion
$\slashl{D}q=0$ 
and $v \cdot D h_{v} = 0$
with $D_{\mu}= \partial_{\mu} - igA_{\mu}$ the covariant derivative.
%
%
We take the vacuum-to-meson matrix element of these identities and
go over to the light-cone limit $x_{\mu} \rightarrow z_{\mu}$.
The terms in the LHS of eqs. (\ref{id1}) and (\ref{id2}) yield  
$\tilde{\phi}_{+}(t)$, $\tilde{\phi}_{-}(t)$ defined above and their derivatives:
$d \tilde{\phi}_{\pm}(t)/dt$ 
and 
$\partial \tilde{\phi}_{\pm}(t, x^{2})/\partial x^{2}|_{x^{2} \rightarrow 0}$,
%
where, via  $\tilde{\phi}_{\pm}(t) \rightarrow \tilde{\phi}_{\pm}(t, x^{2})$,
we extend the definitions in eq.(\ref{phi}) to the case $z \rightarrow x$
($x^{2} \neq 0$).
The last term of (\ref{id2}), the derivative over total translation,
yields contribution with
\be 
  \bar{\Lambda} = M - m_{b} =
 \frac{iv\cdot \partial \langle 0| \bar{q} \Gamma h_{v} |\bar{B}(p) \rangle}
  {\langle 0| \bar{q} \Gamma h_{v} |\bar{B}(p) \rangle}\ .
\label{lambda}
\ee
This is the usual ``effective mass'' of meson states in the 
HQET.\cite{Neubert:1994mb}

The terms given by an integral of quark-antiquark-gluon operator
are expressed by the three-particle distribution amplitudes
corresponding to the higher-Fock components of the meson wave function.
Through the Lorentz decomposition of the three-particle light-cone 
matrix element, we define the three functions 
$\tilde{\Psi}_V \,(t\,,\,u)$, $\tilde{\Psi}_A \,(t\,,\,u)$,
$\tilde{X}_A \,(t\,,\,u)$
as the independent three-particle distributions:\cite{KKQT}
\bea
 \lefteqn{\langle 0 | \bar{q} (z) \, g G_{\mu\nu} (uz)\, z^{\nu}
      \, \Gamma \, h_{v} (0) | \bar{B}(p) \rangle}\nonumber \\  
  &=& \frac{1}{2}\, f_B M \, {\rm Tr}\, \left[ \, \gamma_5\,
      \Gamma \, 
        \frac{1 + \slashs{v}}{2}\, \biggl\{ ( v_{\mu}\slashs{z}
         - t \, \gamma_{\mu} )\  \left( \tilde{\Psi}_A (t,u) 
   - \tilde{\Psi}_V (t,u) \right)
      \right. \nonumber\\
  & & \qquad\qquad\qquad\qquad  - i \, \sigma_{\mu\nu} z^{\nu}\,
           \tilde{\Psi}_V (t,u)
       + \left. \frac{z_{\mu}}{t} \, ( \slashs{z} - t ) \, 
   \tilde{X}_A (t,u)\, \biggr\} \, \right]\ . \label{3elements}
\eea
%

{}From the two identities (\ref{id1}) and (\ref{id2}), we eventually obtain 
the four independent 
differential equations for the distribution amplitudes,\cite{KKQT}
which are exact in QCD in the heavy-quark limit.
This system of four differential equations can be organized
into the two sets, so that the first set of two equations  
does not involve the derivatives with respect to the transverse separation,
$\partial \tilde{\phi}_{\pm}(t, x^{2})/\partial x^{2}|_{x^{2} \rightarrow 0}$, 
while the second set of two equations involves them;
the second set is uninteresting for our purpose.
The first set of equations is given by\cite{KKQT}
%
\bea
 \omega \frac{d \phi_{-}(\omega)}{d \omega}
  &+& \phi_{+}(\omega) = I(\omega)\ ,
  \label{mde1} \\
  \left(\omega - 2 \bar{\Lambda}\right)\phi_{+}(\omega)
 &+& \omega \phi_{-}(\omega) = J(\omega) \ , \label{mde2}
\eea
where we have introduced the 
momentum-space 
distribution amplitudes $\phi_{\pm}(\omega)$ as
%
$\tilde{\phi}_{\pm}(t) = \int d\omega \ e^{-i \omega t}
  \phi_{\pm}(\omega)$
%
with  $\omega v^{+}$ the light-cone projection
of the light-antiquark momentum in the $B$ meson.
$I(\omega)$ and $J(\omega)$ of eqs. (\ref{mde1}) and (\ref{mde2}) denote the 
``source'' terms due to three-particle amplitudes as\cite{KKQT}
\be
 I(\omega) = 2\frac{d}{d\omega}
   \int_{0}^{\omega}d\rho \int_{\omega - \rho}^{\infty}\frac{d\xi}{\xi}
  \frac{\partial}{\partial \xi}\left[ \Psi_{A}(\rho, \xi) 
   - \Psi_{V}(\rho, \xi)\right] \ , \label{si} 
\ee
and similarly for $J(\omega)$.
Here the three-particle amplitudes in the momentum space are defined as
%
$\tilde{F}(t, u) = \int d\omega d \xi \
  e^{-i(\omega  + \xi u)t} F(\omega, \xi)$
with $F=\{ \Psi_{V}, \Psi_{A}, X_{A}\}$.
%

%
A system of equations (\ref{mde1}), (\ref{mde2})
can be solved for $\phi_{+}(\omega)$ and $\phi_{-}(\omega)$
with boundary conditions $\phi_{\pm}(\omega) = 0$
for $\omega < 0$ or $\omega \rightarrow \infty$, and with
normalization condition 
$\int_{0}^{\infty}d\omega \phi_{\pm}(\omega)= \tilde{\phi}_{\pm}(0)= 1$.
The solution can be decomposed into two pieces as
\be
  \phi_{\pm}(\omega) = \phi_{\pm}^{(WW)}(\omega) 
  + \phi_{\pm}^{(g)}(\omega) \ ,
\label{decomp}
\ee
where $\phi_{\pm}^{(WW)}(\omega)$ are the solution of
eqs. (\ref{mde1}) and (\ref{mde2}) with 
$I(\omega)=J(\omega)=0$,
which corresponds to $\Psi_{V}=\Psi_{A}=X_{A}=0$ (``Wandzura-Wilczek approximation'').
$\phi_{\pm}^{(g)}(\omega)$ denote the pieces induced by the source terms.

Eq.(\ref{mde1}) alone, with $I(\omega)=0$, is 
equivalent to a usual Wandzura-Wilczek type relation derived
in Ref.\cite{Beneke:2001wa}:
%
$\phi_{-}^{(WW)}(\omega) = \int_{\omega}^{\infty} 
  d\rho \ \phi_{+}^{(WW)}(\rho)/\rho$.
%
Combining eqs.(\ref{mde1}) and (\ref{mde2}), we are able to obtain 
the analytic solution explicitly as ($\omega \ge 0$)
\be
 \phi_{+}^{(WW)}(\omega) = \frac{\omega}{2 \bar{\Lambda}^{2}} 
 \theta(2 \bar{\Lambda} - \omega) \ , \;\;\;\;\;\;\;\;
 \phi_{-}^{(WW)}(\omega) = 
  \frac{2 \bar{\Lambda} - \omega}{2 \bar{\Lambda}^{2}} 
  \theta(2 \bar{\Lambda} - \omega) \ . \label{solm}
\ee
%
The solution for $\phi_{\pm}^{(g)}$ can be obtained
straightforwardly, and reads ($\omega \ge 0$):
\bea
 \phi_{+}^{(g)}(\omega) &=& \frac{\omega}{2\bar{\Lambda}}\Phi(\omega)\ ,
\;\;\;\;\;\;\;\;\;\;\;\;\;\;\;\;\;\;
  \phi_{-}^{(g)}(\omega) =
  \frac{2\bar{\Lambda}-\omega}{2\bar{\Lambda}}\Phi(\omega)
 + \frac{J(\omega)}{\omega}\ ,\label{solmg}\\
%
%
 \Phi(\omega) &=& \theta(2\bar{\Lambda}-\omega)
 \left\{\int_{0}^{\omega}d\rho \frac{K(\rho)}{2\bar{\Lambda} - \rho}
 -\frac{J(0)}{2\bar{\Lambda}}\right\}
 - \theta(\omega - 2\bar{\Lambda}) \int_{\omega}^{\infty}
  d\rho \frac{K(\rho)}{2\bar{\Lambda} - \rho} \nonumber \\
   &-& \int_{\omega}^{\infty}d\rho
  \left( \frac{K(\rho)}{\rho} + \frac{J(\rho)}{\rho^{2}} \right) \ ,
\label{Phi}
\eea
with
%
$K(\rho) = I(\rho) + \left[ 1/(2\bar{\Lambda}) -
    d/d\rho\right]J(\rho)$.
%
The solution (\ref{decomp}) with eqs.(\ref{solm})-(\ref{Phi})
is exact and 
presents our principal result.

It is also straightforward to derive the Mellin moments
$\langle \omega^{n} \rangle_{\pm} \equiv \int d\omega \ \omega^{n} \phi_{\pm}(\omega)$ 
($n= 0, 1, 2, \cdots$)
of our solution. 
Because the analytic expression for general moment $n$ is somewhat complicated,\cite{KKQT}
we present some examples for a few low moments:
$\langle \omega \rangle_{+} = 4\bar{\Lambda}/3$, 
$\langle \omega \rangle_{-}= 2\bar{\Lambda}/3$, and  
\be
\langle \omega^{2} \rangle_{+} = 2 \bar{\Lambda}^{2}+\frac{2}{3}\lambda_{E}^{2}
 +\frac{1}{3}\lambda_{H}^{2} \ ,\quad\ \;\;\;\; 
 \langle \omega^{2} \rangle_{-} = \frac{2}{3} \bar{\Lambda}^{2}
 +\frac{1}{3}\lambda_{H}^{2} \ ,
\label{mome2}
\ee
where $\lambda_{E}$ and $\lambda_{H}$ 
are due to $\phi_{\pm}^{(g)}$, and 
are related to the chromoelectric and chromomagnetic fields
in the $B$ meson rest frame as
%
$\langle 0 |\bar{q} g \mbox{\boldmath $E$}\cdot\mbox{\boldmath $\alpha$}
 \gamma_{5}h_{v} |\bar{B}(\mbox{\boldmath $p$}=0)\rangle
 = f_{B}M \lambda_{E}^{2}$,
$\langle 0 |\bar{q} g \mbox{\boldmath $H$}\cdot\mbox{\boldmath $\sigma$}
 \gamma_{5}h_{v} |\bar{B}(\mbox{\boldmath $p$}=0)\rangle
 = if_{B}M \lambda_{H}^{2}$.
%
These results for $n=1, 2$ coincide with the relations
obtained by Grozin and Neubert,\cite{Grozin:1997pq}
who have derived their relations by analyzing matrix elements of 
{\it local} operators corresponding to these moments.
Our solution (\ref{decomp}) from nonlocal operators
gives generalization of theirs to $n \ge 3$.


Our results 
reveal that
the leading-twist distribution amplitude $\phi_{+}$ as well as
the higher-twist $\phi_{-}$
contains the three-particle contributions,
which is in contrast with the case of the light mesons.\cite{Braun:1990iv}
We note that
there exists
an estimate of $\lambda_{E}$ and $\lambda_{H}$
by QCD sum rules,\cite{Grozin:1997pq} 
but any estimate of the higher moments is not known.
{}Further investigations are required 
to clarify the effects of multi-particle states.
In this connection, we note that 
the shape of our Wandzura-Wilczek
contributions (\ref{solm}),
which are determined uniquely in analytic form in terms of $\bar{\Lambda}$, 
is rather different from various ``model'' distribution amplitudes that have been used
in the existing literature (see Ref.\cite{Grozin:1997pq} and references therein).


To conclude, our solution 
provides the powerful framework for building up the $B$ meson
light-cone distribution amplitudes and their phenomenological applications,
because the solution is exact and satisfies all relevant QCD constraints.

\bigskip
We 
thank the organizers and all others who helped make
the symposium successful.
We thank T. Onogi for fruitful discussions.
The work of J.K. was supported in part by the Monbu-kagaku-sho Grant-in-Aid
for Scientific Research No.C-13640289.
The work of C-F.Q. was supported by the Grant-in-Aid of JSPS committee.

\end{document}